# Model Simplification Through Refinement

Dmitry Brodsky
Department of Computer Science
University of British Columbia

Benjamin Watson
Department of Computing Science
University of Alberta

*Abstract*

As modeling and visualization applications proliferate, there arises a need to simplify large polygonal models at interactive rates. Unfortunately existing polygon mesh simplification algorithms are not well suited for this task because they are either too slow (requiring the simplified model to be pre-computed) or produce models that are too poor in quality. These algorithms are also not able to handle extremely large models.

We present an algorithm suitable for simplification at interactive speeds and of extremely large models. The algorithm is fast and can guarantee displayable results within a given time limit. Results also have good quality. Inspired by splitting algorithms from vector quantization literature, we simplify models in reverse, from coarse to fine. Approximations of surface curvature guide the simplification process. Previously produced simplifications can be further refined by using them as input to the algorithm.

## 1 Introduction

Many of today's applications require simplification of polygonal models at interactive speeds. Modeling applications must simplify and display extremely large models at interactive rates. In visualization applications isosurfaces from high dimensional data sets must be computed, simplified, and rendered in close to real-time. In dynamically modifiable virtual environments, newly generated surfaces are typically over-tessellated and must be simplified for display at interactive speeds. As the size of polygonal models balloons simplification algorithms have to scale to gracefully handle these extremely large models.

An ideal simplification algorithm that is able to simplify at interactive rates and handle extremely large models would possess several characteristics. Most importantly, the algorithm must guarantee displayable results within a specified time limit. Second, the algorithm must provide good control of output model size if results are to be displayable. It is also very important that the output model quality remains reasonable, despite stringent time constraints. If time demands require the output of a crude simplification, then the algorithm should allow for later refinement of that output. Finally, for interactive display it would be useful if the algorithm produced a continuous level of detail hierarchy instead of several discrete levels of detail.

Most existing simplification algorithms are far too slow to be used in interactive applications. Some vertex clustering algorithms [14, 18] are very fast, but control of output quality and size is quite poor. Moreover this output is difficult to refine and to organize into a continuous level of detail hierarchy.

Our algorithm, *R-Simp*, was inspired by splitting algorithms from the vector quantization literature [6]. The algorithm simplifies in reverse from coarse to fine, allowing us to guarantee a displayable result within a specified time limit. At every iteration of the algorithm, the number of vertices in the simplified model is known, enabling control of output model size. We use curvature to guide the simplification process, permitting preservation of important model features, and thus a reasonable level of output model quality. Performing simplification in a reverse direction makes it possible to refine intermediate output as long as some state information is saved. Finally with its divide and conquer approach, R-Simp can easily be extended to create continuous level of detail hierarchies. R-Simp's complexity is $O(n_i \log n_o)$, where $n_i$ is the size of the input model and $n_o$ is the size of the output model. This enables R-Simp to scale linearly with respect to input size for a given output size. With all these traits, R-Simp is well suited for simplification in interactive environments.

We also look to vector quantization to form a taxonomy of existing simplification algorithms. In sections 2, 3 and 4 we review vector quantization, related research, and curvature. The details of the algorithm are discussed in section 5. In section 6 we examine the performance of the algorithm and compare it to QSlim [5] and a vertex clustering algorithm [18]. Sections 7 and 8 present other possible applications of R-Simp and conclusions.

## 2 Vector quantization

Vector quantization (VQ) is the process of mapping a vector in a large set $S \subset \mathbb{R}^n$ into a smaller set $C \subset \mathbb{R}^n$. More precisely, a quantizer is a function $Q : \mathbb{R}^n \to C$ where

$\mathsf{C} = \{\vec{v}_i \in \mathbb{R}^n | 1 \leq i \leq N\}$. $\mathsf{C}$ is called the *codebook*. The challenge is finding $\mathsf{C}$ such that it optimally represents all vectors in $S \subset \mathbb{R}^n$. The codebook $\mathsf{C}$ partitions the set $\mathsf{S}$, since each $\vec{v}_i$ represents multiple vectors from $\mathsf{S}$. A single partition of $\mathsf{S}$ is called a *cell* and $\vec{v}_i$ is the *centroid* of the cell. The difference between a vector $\vec{v}_i$ and an input vector $\vec{u}$ is called *distortion*. When the distortion for an input vector $\vec{u}$ is minimal for all $\vec{u} \in \mathsf{S}$ then the codebook is called optimal.

In [6], Gersho and Gray present four basic types of VQ algorithms. Using these four types we will create a taxonomy of existing simplification algorithms.

*Product code* algorithms use scalar quantizers that are independently applied to each input vector element.

In *pruning* algorithms, the codebook initially contains all the vectors in the input set $\mathsf{S}$. The codebook entry that increases distortion least is removed; removals continue until the desired codebook size is reached. Alternatively, the codebook is initially empty, and each of the input vectors is considered in succession. If representing any vector with the current codebook would result in distortion over a given threshold, the vector is added to the codebook.

*Pairwise nearest neighbor* algorithms also set the initial codebook to contain all the vectors in $\mathsf{S}$. All possible cell pairs are considered and the pair that introduces the least distortion is merged. Merging continues until the desired codebook size or distortion tolerance is reached.

In *Splitting* algorithms, the codebook initially contains a single cell. The cell with the most distortion is located and then split. Splitting continues until the required distortion or codebook size is reached.

## 3 Vector Quantization and Simplification

Simplification relates to quantization as follows: a centroid equates to a primitive (vertex, line, or polygon) or a set of primitives in the simplified model. For most vertex merge algorithms, the centroid is a single vertex and associated faces. A cell equates to a set, cluster, of faces in the original model. There are a few ways in which model simplification differs from vector quantization. For example in model simplification, two disjoint faces do not make up an ideal cluster, while in image quantization a cluster with two separate pixels is perfectly acceptable.

Rossignac and Borrel [18] proposed an algorithm that applies a product codes technique to the model vertices. Cells are formed with a uniform voxelization; the centroid is usually chosen as the mean of the vertices in each cell (weighted averages or maxima are common alternatives). Only a linear pass through the vertices is required to simplify the surface. The result is an extremely fast algorithm that produces poor simplifications. He et al [7] proposed a similar and slower algorithm that makes use of a low pass three dimensional filter. Low and Tan [14] developed a vertex clustering algorithm that non-uniformly subdivided the model's volume. Cells are centred on the most important vertices in the model.

The simplification algorithms taking the pruning approach are generally not as fast as the product code algorithms, but they produce better simplifications. Two such algorithms [8, 12] work by growing coplanar patches. When a face cannot be added to a patch without violating a co-planarity threshold, it is re-triangulated with fewer polygons and added to the codebook. Other algorithms [19, 20] work by removing or pruning away single vertices. The algorithm described in [19] simply removes a vertex whose surrounding faces are relatively coplanar and re-triangulates the created hole, while the algorithm described in [20] adds a completely new set of vertices and tries to prune away as many of the old vertices as possible.

There are many algorithms, commonly called vertex merge or edge collapse algorithms, that use the pairwise nearest neighbour approach [1, 3, 5, 9, 10, 13, 17]. These algorithms tend to produce the best simplification results but are often quite slow. The algorithms assign weights to each vertex merge and use a priority queue to locate the merge with minimum cost. They merge the vertices (merge the cells), recompute the affected vertex pairs, and iterate. The algorithms continue until the required model size or error tolerance is reached. The algorithms differ in how they assign weights to a vertex merge and how they determine the location of merged vertices (calculate centroids).

To our knowledge R-Simp is the only simplification algorithm based on the splitting technique. In R-Simp we treat simplification as quantization of face normals as opposed to colour ($x, y, z$ instead of $R, G, B$). Our goal when splitting is to create cells containing the most planar surface possible (the variation in face normals is small). Thus, cells that contain little curvature are split less than cells that contain more curvature.

## 4 Curvature

One common measure of surface curvature is called *normal curvature* [15]. Normal curvature is the rate of change of the normal vector field $U$ on a surface $S$ in direction $\vec{u}$, where $\vec{u}$ is a unit vector tangent to the surface $S$ at point $p$. There are two important normal curvature extrema called *principle curvatures*, these are the maximum ($k_1$) and minimum ($k_2$) values of normal curvature. The directions corresponding to these principle curvatures are called *principle directions*.

Since these curvature measures are defined for in-finitely small patches, they provide a good description of the local surface around a point. However, they do not work well for larger surface patches with multiple scales of curvature. (e.g. asphalt looks flat from a distance but can feel quite rough close up). R-Simp requires measures of orientation change, curvature, for large patches. We will use the term *normal variation* to refer to orientation change in large patches.

## 5 The R-Simp algorithm

Unlike other algorithms, R-Simp starts with a coarse approximation of the model and refines it until the desired model complexity is reached. The algorithm begins with the triangulated model in a single cluster (a cluster is a collection of faces from the original model). The initial cluster is then subdivided into eight sub-clusters. These eight sub-clusters are then iteratively subdivided until the required number of clusters (vertices) is reached. Clusters are chosen for subdivision based on the amount of normal variation on the surface in the cluster.

The R-Simp algorithm can be broken down into three stages.

- **Initialization:** In this stage we create global face (gfl) and vertex (gvl) lists, as well as vertex-vertex and vertex-face adjacency lists. We also create the eight initial clusters.

- **Simplification:** In this stage the model is simplified. The simplification consists of four steps:

    1. Choose the cluster that has the most face normal variation.
    2. Partition (split) the cluster based on the amount and direction of the face normal variation.
    3. Compute the amount of face normal variation in each of the sub-clusters.
    4. Iterate until the required number of clusters (vertices) is reached.

- **Post Processing:** For each cluster that is left, compute a representative vertex (centroid). Retriangulate the model.

### 5.1 Data structures

The principle data structure in this algorithm is the Cluster. It stores all the information necessary to determine face normal variation and to compute the representative vertex. It contains two arrays of indices, for vertices (vl) and faces (fl), that index into two global lists of the vertices and faces from the original model (gfl and gvl). The Cluster also contains the mean normal ($\vec{mn}$) that is the area-weighted mean of all the face normals in the cluster and is computed by Equation 1.

$$\vec{mn} = \sum_i^N \vec{n}_i a_i \quad (1)$$

where $N$ is the number of faces in the cluster, $\vec{n}_i$ is the normal of face $i$, and $a_i$ is the area of face $i$. The Cluster also holds the mean vertex ($mv$) for the cluster, the amount of normal variation ($nv$), and the total area of the faces in the cluster.

Two other important data structures are the Face and the Vertex data structures which make up gfl and gvl respectively. The Face contains a list of vertices that make up the face, its normal, the face area, and its midpoint. The Vertex contains adjacency information for all the vertices and faces adjacent to it.

The vertices in the Face data structure are indices into gvl. The adjacency lists for the faces and the vertices in the Vertex data structure are also indices into gfl and gvl.

### 5.2 Initialization

During the initialization stage gfl and gvl are constructed and the initial eight clusters are created. The initial clusters are created by partitioning the model using three axis aligned planes that are positioned in the middle of the model's bounding box. We then compute the amount of face normal variation in each of these clusters (see Section 5.3). These eight clusters are then inserted into a priority queue sorted by the amount of face normal variation.

### 5.3 Choosing the cluster to partition

In the simplification stage of our algorithm the first step is to choose a cluster in which the face normals vary the most (the cluster at the head of the queue). We compute the amount of face normal variation using the area-weighted mean ($\vec{mn}$) of the face normals.

The flatter the surface, the larger the magnitude of $\vec{mn}$. If all the faces are coplanar, the magnitude of $\vec{mn}$ will equal the area of the surface in the cluster. We define this component $cp$ of our face normal variation measure as follows:

$$cp = \frac{||\vec{mn}||}{\sum_i^N a_i} \quad (2)$$

Even if the surface in a cluster is extremely small it can contain a large amount of curvature. In order to prevent small, highly curved details (e.g. a small spring in an engine) from dominating the simplification we must make our normal variation measure ($nv$) sensitive to size. To do this, we scale $cp$ by the ratio of the surface area in the

cluster to the model surface area:

$$nv = \frac{\sum_i^N a_i}{\sum_i^M a_i}(1 - cp) \quad (3)$$

where $M$ is the number of faces in the model. We complement $cp$ so that $nv$ increases as face normal variation increases. In the remainder of this paper the term "normal variation" refers to variation of face normals.

### 5.4 Describing the pattern of normal variation

The next step is to describe normal variation in the chosen cluster. We follow Gersho and Gray [6] who suggest principle component analysis (PCA) [11] as a way of determining how to split cells when using a splitting algorithm. In PCA a covariance matrix is formed from the data set of interest. The eigenvectors of this matrix are aligned according to the pattern of variation in the data set. Garland [4] showed that if the covariance matrix is formed with normal vectors, the eigenvectors are generally related to the principal directions of normal curvature. Specifically, the largest eigenvalue and corresponding eigenvector represent the mean normal of the surface. Usually the second and third largest eigenvalues and corresponding eigenvectors represent the directions of maximum and minimum curvature.

The covariance matrix $\mathcal{A}$ around the mean $[0, 0, 0]$ is defined by:

$$\mathcal{A} = \sum_i^N \vec{n}_i \vec{n}_i^T \quad (4)$$

We compute the eigenvalues and eigenvectors using the Jacobi method [16].

### 5.5 Partitioning the cluster

Partitioning the cluster consists of four steps. First, we must determine how many planes to use to partition the cluster. Second, we must orient the planes. Finally, we must position the planes and create new sub-clusters.

A cluster is partitioned into two, four, or eight sub-clusters depending on the amount of curvature. Let $c_{mn}$, $c_M$ and $c_m$ equal the eigenvalues in descending order (the second and third largest eigenvalues relate to $k_1$ and $k_2$, the magnitudes of principle curvature). Let $\vec{c}_M$ and $\vec{c}_m$ represent the corresponding eigenvectors (these are related to the directions of maximum and minimum curvature).

If all eigenvalues are of similar magnitude the pattern of normal variation is unclear. We test for this by comparing the eigenvalues as follows: both $c_M < 2c_m$ and $c_{mn} < 2c_M$ must be true. In this case we partition the cluster into eight sub-clusters. One partitioning plane is perpendicular to $\vec{c}_M$, the second plane is perpendicular to $\vec{c}_m$, and the third plane is perpendicular to $\vec{mn}$.

Otherwise, if $\frac{c_M}{c_m} <= 4$ then the surface is most likely hemispherical since there is significant curvature in both the minimum and maximum directions of curvature. In this case we partition the cluster into four sub-clusters. One partitioning plane is perpendicular to $\vec{c}_M$ and the other plane is perpendicular to $\vec{c}_m$.

In all remaining cases $\frac{c_M}{c_m} > 4$ and the surface is most likely cylindrical since most of the curvature is in one direction. In this case we partition the cluster into two sub-clusters. The partitioning plane is perpendicular to $\vec{c}_M$.

We must now position the partitioning planes in the cluster. Ideally the surface should be partitioned along any ridges or through any elliptical bumps. However, locating such features is difficult, instead we do the following: first we compute the vector $\vec{c}_{M\perp}$, which is the projection of $\vec{c}_M$ onto $P_{mn}$, the plane defined by $\vec{mn}$ and the cluster's mean vertex ($mv$). We then project the midpoint of all the faces in the cluster onto $P_{mn}$ and find the mean of all projected midpoints that fall within 2.5 degrees of $\vec{c}_{M\perp}$. The resulting point is the position for the partitioning plane(s).

Sub-clusters are created by partitioning the vertices in a cluster. The membership of a vertex depends on which side of the partitioning plane(s) it falls on. The faces follow the vertices to the sub-clusters. A face may belong to two or three clusters if the vertices of the face fall into different sub-clusters.

Even if the entire model is topologically connected, a given cluster may contain two or more disconnected components. Approximating these components with a single vertex can introduce severe distortion. We have found it useful to perform a topology check to determine if a new cluster contains topologically disjoint components.

The topology check is a breadth first search on the vertices and edges contained in a cluster. We use a bit array to record the vertices visited during the search, making it linear in complexity. If the cluster contains disjoint components, each component is placed into a separate cluster. Although this topology check increases the overall simplification time, the resulting increase in quality of the simplification is considerable.

### 5.6 Post processing

Once the simplification stage is finished two tasks remain. The first is to compute the location of the representative vertex ($v$) for each cluster. The second is to re-triangulate the output surface.

To represent a cluster's faces as accurately as possible, $v$ should be as close as possible to all the faces. [5, 13, 17] all minimize the summed distance from the planes containing the cluster's faces. [5, 13] minimize the squared

distance:

$$Q(v) = v^T \mathcal{A} v + 2\mathcal{B}^T v + c \quad (5)$$

Where $n_i + d_i = 0$ is the plane equation for face $i$, $\mathcal{A}$ is as previously defined, $\mathcal{B} = \sum_i^N d_i \vec{n}_i$, and $c = \sum_i^N d_i^2$.

Since $Q$ is a quadratic then $Q(v)$ is minimum when its partial derivatives equal zero. This occurs when:

$$v_{min} = -\mathcal{A}^{-1}\mathcal{B} \quad (6)$$

We re-triangulate using a method similar to that used by [18]. After $v_j$ is computed for each cluster, the $v_j$s are output to a simplified vertex list `svl`. In `gvl`, all vertex references (`rvp`) contained in the cluster $j$ are pointed at the new entry in `svl`. We then traverse the global face list `gfl`. Any face referencing three different vertices in `svl` is retained and output to the simplified face list `sfl`. All other faces have degenerated into lines or points and are discarded.

## 6 Results

Simplification algorithms are usually judged by two criteria. The first criterion is speed, the time required to simplify a model. The second and more difficult to measure criterion is quality. Intuitively speaking, quality of a simplification is its appearance or its geometric accuracy.

In the following subsection we present execution times for two different input models. We also present quality results, including images allowing for comparison of appearance and geometric accuracy measured with the *Metro* [2] tool.

### 6.1 Performance

Five different models were used in our comparisons. All models were simplified on a 195 MHz R10000 SGI Onyx2 with 512 MB of main memory.

We compared R-Simp to two other simplification algorithms. We chose the fastest vertex clustering algorithm and the fastest vertex merge algorithm. The first is Rossignac and Borrel's [18] vertex clustering algorithm (with unweighted centroid calculations). The second, QSlim [5], is one of the fastest vertex merge algorithms.

Figure 1 compares the performance of R-Simp to QSlim and vertex clustering with the Stanford bunny. R-Simp is considerably faster than QSlim; it is able to produce a simplified model of up to 20000 polygons before QSlim removes a single face. R-Simp's complexity is $O(n_i \log n_o)$ where $n_i$ is the input model size and $n_o$ is the output model size. Thus R-Simp is linear for a fixed output size. The speed of vertex clustering is not related to output size.

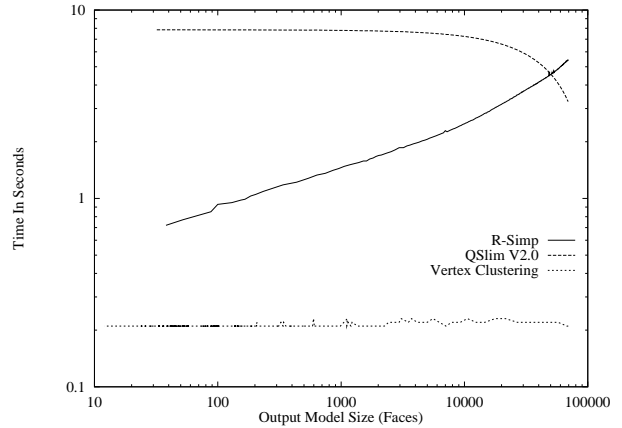

Figure 1: *The effect of output model size on simplification time for the Stanford bunny.*

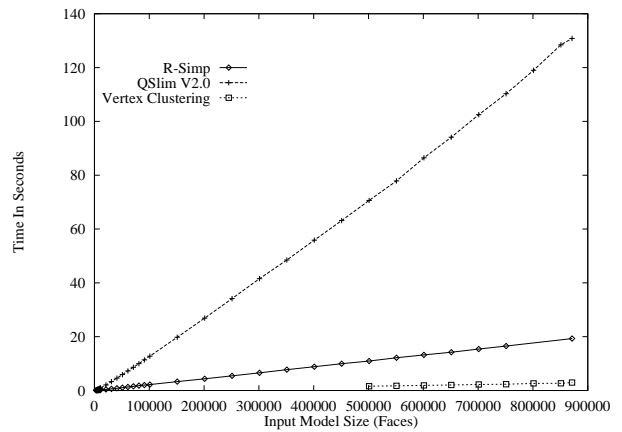

Figure 2: *The effects of input model size on simplification time. Output model size is 2100 polygons.*

Figure 2 shows how the size of the input model affects simplification time. The dragon was initially simplified using QSlim to various sizes. These models were then simplified by R-Simp, QSlim, and vertex clustering to 2100 polygons. As the graph shows, the larger the input model, the longer it takes to simplify. However, QSlim's curve is significantly steeper than R-Simp's. Vertex clustering is fastest but is affected by input size.

To compare model quality we took seven models and simplified them. Table 1 summarizes the results. The table shows the mean Hausdorff distance between the original and the simplified surface as a percentage of the diagonal of the bounding box of the original surface [2].

Figures 3a-h show the original bunny and dragon models and the corresponding simplifications produced by all three algorithms.

| Model | Input # Faces | Output # Faces | % of Error & Simplification Time (s) | | | | | |
|---|---|---|---|---|---|---|---|---|
| | | | Cluster | | RSimp | | QSlim | |
| Bunny | 69451 | 1600 | 0.302% | 0.09 | 0.155% | 1.58 | 0.071% | 7.85 |
| Cow | 5782 | 1600 | 0.256% | 0.03 | 0.118% | 0.12 | 0.060% | 0.45 |
| Dragon | 871306 | 2100 | 0.428% | 0.29 | 0.241% | 18.9 | 0.175% | 129 |
| Horse | 96966 | 1600 | 0.266% | 0.05 | 0.147% | 2.29 | 0.052% | 11.5 |
| Chair | 2481 | 800 | 0.658% | 0.00 | 0.215% | 0.05 | 0.019% | 0.17 |
| Torus | 20000 | 400 | 0.460% | 0.05 | 0.265% | 0.32 | 0.160% | 1.74 |
| Spring | 9386 | 800 | 1.012% | N/A | 0.594% | 0.13 | 0.295% | 0.75 |
| Mean | | | 0.483% | 0.09 | 0.242% | 3.34 | 0.119% | 21.6 |

Table 1: *Simplification error and time of RSimp, QSlim, and vertex clustering. The error is the percentage of mean error returned by Metro [2].*

## 6.2 Discussion

As we noted earlier, in applications where the model is created in response to user input there is no time for pre-computation. Models must be simplified in reasonable time. Generation of iso-surfaces is one example of such an application; the surface is not known before a user inputs an iso-value. The models in dynamically modifiable virtual environments are by definition not pre-computable. In many modeling applications users are constantly modifying complex models. Applications that deal with extremely large models must ensure that the simplification algorithm is able to handle models of arbitrary size.

Algorithms useful for such applications should possess several characteristics:

- *Interactive response*: Most importantly, algorithms should be able to guarantee displayable results within a specified time limit. R-Simp's speed and coarse to fine pattern of simplification make it ideal for this application. Vertex clustering is even faster although precise control of execution time is difficult. Because QSlim is slower and simplifies from fine to coarse it cannot make any time guarantees.

- *Control of output model size*: Control of output model size is important if results are to be displayable. R-Simp and QSlim provide a straight forward way to control the output model size but most vertex clustering algorithms do not. In these algorithms one can only specify the number of voxels; the number of vertices and faces will typically be much smaller. Thus, if displayability is to be guaranteed, quality suffers.

- *High output quality*: Algorithms should output models of the best possible quality despite time constraints. QSlim clearly generates the best quality models. R-Simp's output quality is not as good, vertex clustering is worst.

- *Iterative improvement*: When time constraints require the output of a crude simplification, it should be possible to refine the result after the time demands have been met. R-Simp's coarse to fine pattern of simplification makes this fairly simple; one must only save the priority queue of clusters. With QSlim there is no need for refinement, the more important question is whether the time constraints could be met. Vertex clustering uses a one pass, one resolution approach and thus refinement is not possible.

- *Continuous level of detail*: Many interactive applications require level of detail hierarchies. Both R-Simp and QSlim are able to produce hierarchies but much of the hierarchy initially output by QSlim will not be displayable because they will be too large to display. Vertex clustering cannot produce hierarchies without fundamental changes to the algorithm.

- *Scalability*: Simplification algorithms need to scale so that they are able to gracefully handle extremely large models. QSlim's complexity is $O(n \log n)$ while R-Simp's is linear for a fixed output size.

To summarize, QSlim generates the best quality models but it is not suitable for interactive applications because it is too slow and cannot guarantee a bounded runtime. Vertex clustering algorithms are extremely fast but generate poor quality models and do not provide an easy way to control output model size. We believe R-Simp's time guarantees and quality/speed tradeoff make it ideal for use in interactive applications.

R-Simp can simplify any model, regardless of topology or manifold characteristics. In output it can simplify topology and thus does not guarantee topology preservation.

## 7 Future work and other applications

The quality of R-Simp's simplifications might be improved by adding a look-ahead feature, comparing the normal variation before and after the cluster split. It should also be possible to modify R-Simp to consider boundaries as well as face and vertex attributes (e.g. colour) during simplification. For large environments consisting of many objects, it should be possible to add a distance threshold to the topology check, so that disjoint but neighbouring objects remain in the same cluster and are merged.

To enable management of the quality/speed tradeoff, R-Simp might be used as part of a two stage simplification process. If speed is particularly important, vertex clustering could be used to simplify the model to a medium level of complexity and the result input to R-Simp. If quality is important, the output of R-Simp could be input to QSlim.

We have already discussed the use of R-Simp for level of detail hierarchies. The bounding box around the clusters in these hierarchies can be used to speed up collision detection. We have experimented with such bounding boxes as an error measure during simplification and found no loss in quality or speed. For view based level of detail control, the error measure should limit the distance between the simplified and original surfaces.

Since R-Simp simplifies in a coarse to fine direction, it should be well suited for application in progressive transmission of 3D models. Approximations of previously uncompressed models could be transmitted quickly.

## 8 Conclusion

We presented R-Simp, an algorithm that simplifies 3D models in reverse and is well suited for interactive applications such as generation of iso-surfaces. Given a limited amount of time most other algorithms cannot guarantee displayable results or results of reasonable quality. R-Simp also allows iterative improvements, precise control of output size, and construction of level of detail hierarchies.

**Acknowledgments**

We would like to thank Oleg Verevka for suggesting the comparison of quantization to model simplification and Carolina Diaz-Goano for her mathematical assistance. We are grateful to Alex Brodsky for all his helpful and insightful comments and to Greg Turk for his comments, geometry filters, and models. This research was supported by an NSERC grant: RGPIN203262-98.

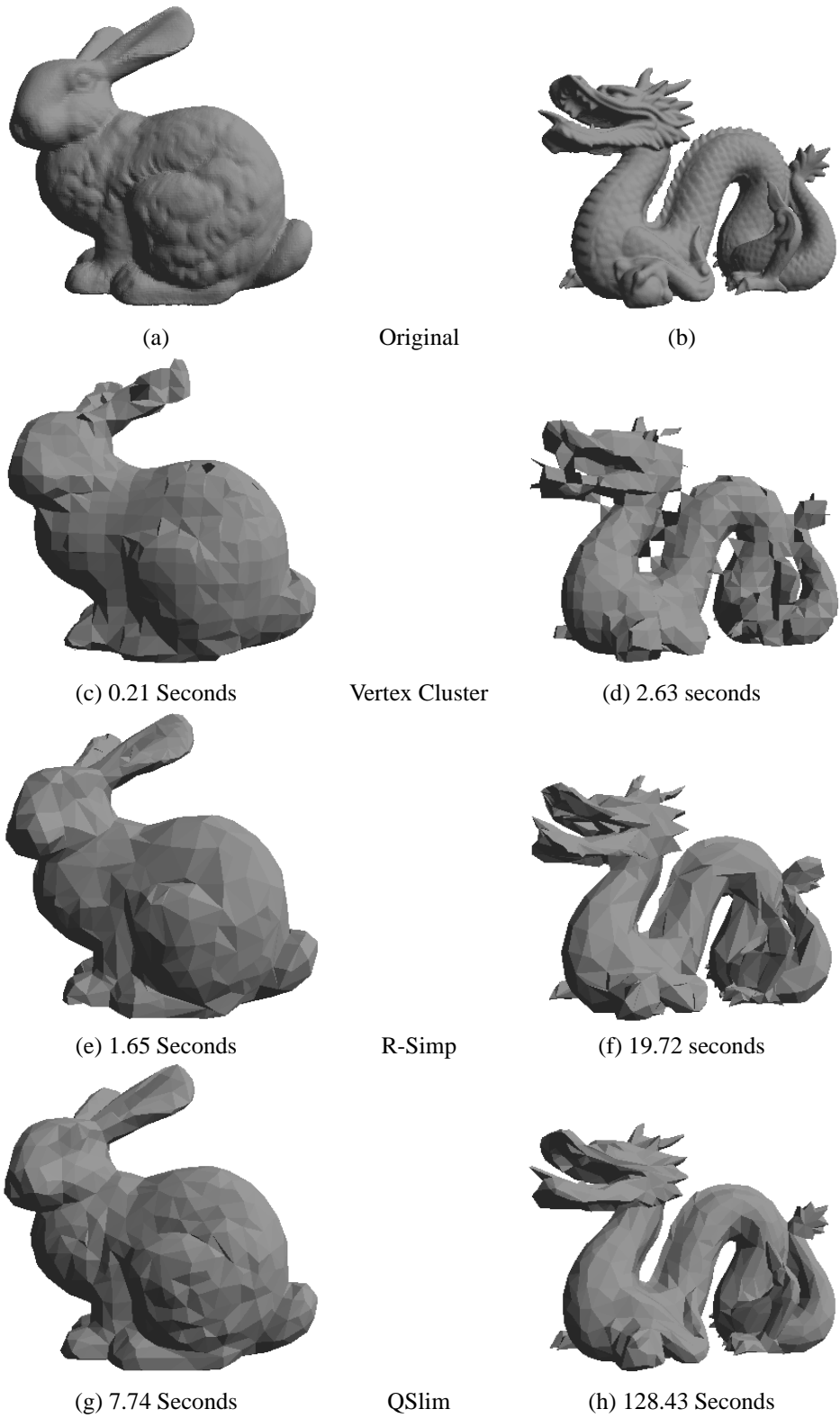

Figure 3: Visual results of the three simplification algorithms. (a) Original bunny 69451 faces. (b) Original dragon 871306 faces. (c)(e)(g) are 1900 faces. (d)(f)(h) are 2500 faces.